\documentclass[twocolumn,floatfix,aps,superscriptaddress]{revtex4}
\usepackage{graphicx}
\usepackage{amsmath}
\usepackage{amssymb}
\usepackage{bm}
\usepackage{hyperref}
\hypersetup{colorlinks=true,urlcolor=blue}

\begin{document}

\title{A road to reality with topological superconductors\footnote{Contribution to a focus issue on topological matter in: Nature Physics \textbf{12}, 618--621 (2016) \href{http://dx.doi.org/10.1038/nphys3778}{DOI:10.1038/nphys3778}}}
\author{Carlo Beenakker and Leo Kouwenhoven}
\begin{abstract}
Topological states of matter are a source of low-energy quasiparticles, bound to a defect or propagating along the surface. In a superconductor these are Majorana fermions, described by a real rather than a complex wave function. The absence of complex phase factors promises protection against decoherence in quantum computations based on topological superconductivity.
\end{abstract}
\maketitle

Quantum mechanics is complex. The only way Erwin Schr\"{o}dinger could get his equation $i\hbar\,d\psi/dt=H\psi$ to work, was to multiply the time derivative of the wave function $\psi$ by the imaginary unit $i$. He complained about that in a 1926 letter \cite{quote1}: ``What is unpleasant here, and indeed directly to be objected to, is the use of complex numbers --- $\psi$ is surely fundamentally a real function.'' But the $i$ was there to stay. Freeman Dyson called this apparently illogical step ``one of the most profound jokes of nature'' \cite{quote2}: ``Schr\"{o}dinger put the square root of minus one into the equation, and suddenly it made sense. Suddenly it became a wave equation instead of a heat conduction equation \ldots And that square root of minus one means that nature works with complex numbers and not with real numbers.''

Topological superconductivity provides a road to reality. Topological superconductors and topological insulators both combine a gapped bulk with gapless surface excitations, governed by a relativistic wave equation. But while the wave function $\psi$ is complex in an insulator, $\psi$ is real in a topological superconductor. A real $\psi$ means that scattering phase shifts are limited to $\pm 1$, which profoundly changes the way quantum interference operates and promises a robustness of phase coherence that a complex $\psi$ lacks.\medskip

\noindent
\textbf{Bogoliubov meets Majorana}\\
The mathematics that allows for a real wave function is simple: If an electron at energy $E$ has a time dependent $\psi\propto e^{-iEt/\hbar}$, then its antiparticle (a ``hole'') has $\psi\propto e^{+iEt/\hbar}$ and a linear superposition would give a real $\psi$. In a physical system this superposition is produced by transitions between states of charge $+e$ and $-e$ that are normally forbidden by charge conservation. This is where the superconductor enters, by providing a reservoir of charge-$2e$ Cooper pairs that absorbs the charge difference. The electron-hole superposition, a socalled Bogoliubov quasiparticle \cite{Bog58}, has a Hamiltonian $H=iA$ that turns out to be purely imaginary. The $i$ then cancels with the $i$ in front of $d\psi/dt$, producing a purely real wave equation $\hbar\, d\psi/dt=A\psi$.

This applies to any superconductor, but in a typical situation the physical consequences of a real $\psi$ remain hidden because of a conspiracy of symmetries: Particle-hole symmetry together with spin-rotation symmetry pairs up the Bogoliubov quasiparticles, so that they are effectively represented by a complex wave function --- in much the same way that a complex number is represented by its real and imaginary parts. While particle-hole symmetry is unavoidable in a superconductor, spin-rotation symmetry can be broken, allowing for an unpaired Bogoliubov quasiparticle with a manifestly real $\psi$. The theoretical physicists who studied this scenario at the turn of the century called it a Majorana fermion \cite{Vol99,Sen00,Rea00,Kit01}, in reference to a hypothetical elementary particle from the early days of quantum physics \cite{Maj37}. The theoretical models that produced Majorana fermions had an exotic superconducting order, with spin-triplet Cooper pairs in a chiral \textit{p}-wave orbital state \cite{Kal15}. (``Chirality'' refers to the $p_x\pm ip_y$  structure of the order parameter.) These were among the first appearances on paper of topological superconductors, but it would take another decade for a breakthrough in our thinking how they might be realized in the laboratory.\medskip

\noindent
\textbf{Routes to topological superconductivity}\\
There may well be materials that develop topological superconductivity on their own, Sr$_2$RuO$_4$ is one long-standing candidate for spin-triplet pairing \cite{Mac03}. However, spin-singlet pairing is overwhelmingly more common. The breakthrough that has opened up a great variety of routes to topological superconductivity is the realization that one can start from a conventional spin-singlet superconductor and use the proximity effect to induce a topologically nontrivial superconducting state in a material with strong spin-orbit coupling \cite{Fu08,Sat09,Sau10,Ali10}.

The reasoning behind such hybrid approaches is that the mechanism that produces a topologically nontrivial state is the same for insulators and superconductors: an inversion of the excitation gap in the bulk that leaves behind a gapless surface state. Quite generally, a gap closing followed by a reopening will invert the sign of the gap and transform a topologically trivial state into a nontrivial state. So to create a topological superconductor we need two competing actors, a bad actor which seeks to kill the induced superconductivity, and a good actor which tries to revive it.

In several implementations \cite{Mou12,Das12,Den12,Alb16}, an InSb or InAs nanowire is covered by a Nb or Al superconductor. The proximity effect pairs electrons of opposite spin in the nanowire, producing a superconducting gap at the Fermi level. A magnetic field tends to align the electron spins and close the gap, while spin-orbit coupling counteracts the alignment and reopens it. This competition creates regions in parameter space where the gap is inverted. Since the nanowire is effectively a one-dimensional (1D) system, the ``surface'' is limited to the end points, where a gapless Majorana state is predicted to appear \cite{Lut10,Ore10}. In an alternative 1D implementation \cite{Nad14} the semiconductor nanowire is replaced by a chain of Fe atoms on a Pb substrate. In such a system the atomic magnetization can play both the roles of the bad and the good actor \cite{Cho11}: aligning the spins locally while disrupting the alignment by a rotation of the magnetic moment from one atom to the next. 

These implementations have in common that the gap inversion is tuned by variation of some parameter (typically the magnetic field or electron density). An alternative route to a 1D or 2D topological superconductor starts from the inverted band gap of a 2D or 3D topological insulator and induces superconductivity in the edge or surface states \cite{Fu08,Fu09}. Experiments in this direction are reported in Refs.\ \onlinecite{Kne12,Vel12,Har14,Pri15}.

No single implementation has yet emerged as the ``ideal'' platform for the study of topological superconductivity, but the great variety of options holds promise for rapid experimental developments. In what follows we give an overview of some of the manifestations of a real Majorana wave function that are waiting to be observed.\medskip

\noindent
\textbf{Majorana metal}\\
While a superconductor is a perfect conductor of electricity, it is typically a poor thermal conductor. In a normal metal the addition of disorder would only make things worse, but a 2D topological superconductor will start to conduct heat if enough defects are introduced \cite{Sen00}. This unusual state of matter is called a ``thermal metal'' or ``Majorana metal'', because Majorana fermions bound to defects are responsible for the heat conduction. 

\begin{figure}[tb]
\centerline{\includegraphics[width=0.8\linewidth]{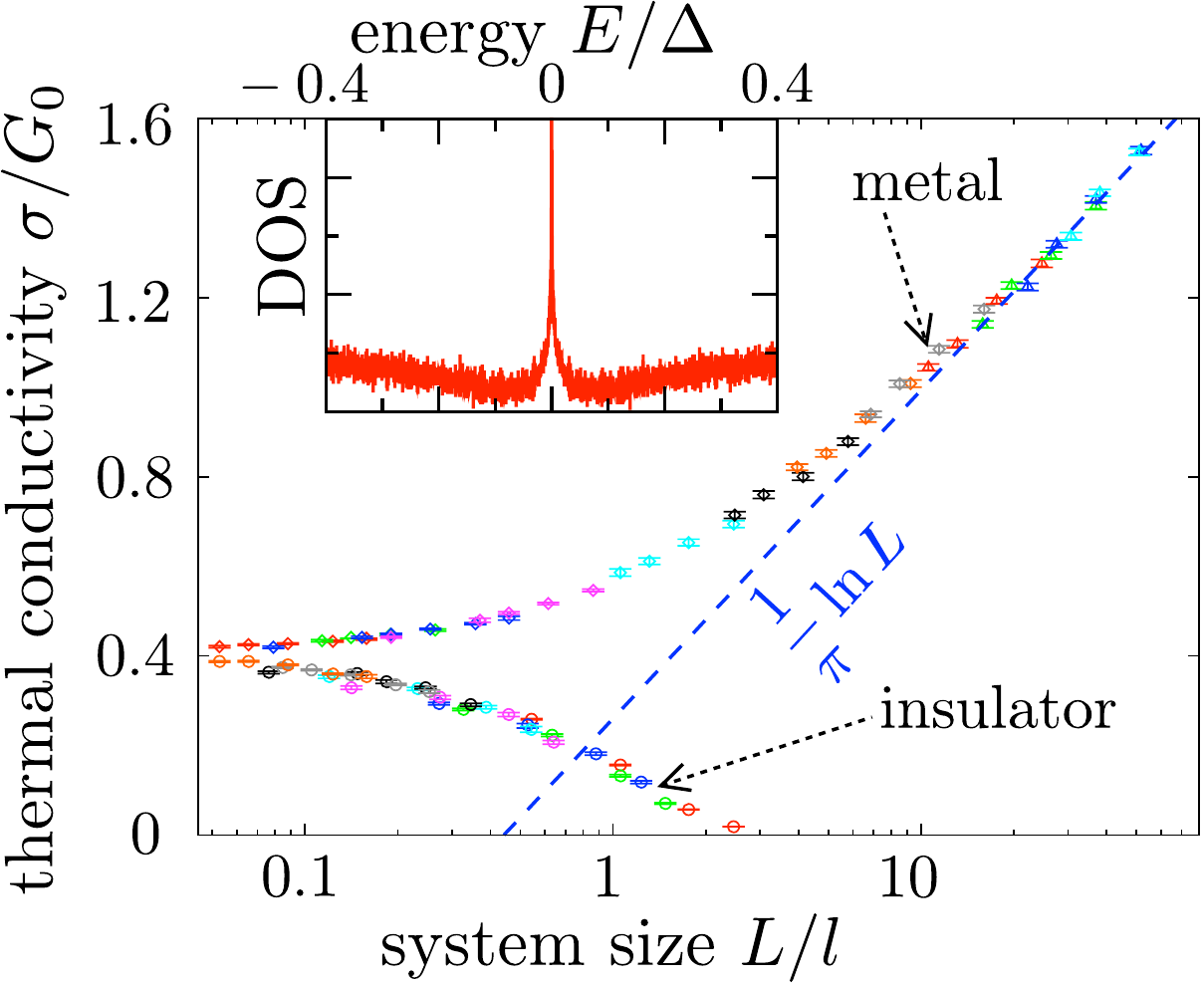}}
\caption{Majorana metal in a computer simulation of a chiral \textit{p}-wave superconductor. The main plot shows the thermal conductivity $\sigma$ as a function of system size $L$ (in units of the mean free path $l$). The data points at different disorder strengths all collapse onto a pair of scaling curves, designated ``metal'' and ``insulator''. The $\ln L$ scaling is characteristic of a Majorana metal \cite{Sen00}, originating from a proliferation of Majorana bound states at $E=0$. The inset shows the corresponding midgap peak in the density of states (DOS). Inset from Ref.\ \onlinecite{Wim10}, main plot from Ref.\ \onlinecite{Med11}.
}
\label{fig_sigmaMI}
\end{figure}

Defects create bound states within the superconducting gap. In a conventional superconductor these will only rarely align in energy $E$ (measured relative to the middle of the gap), so they are not an effective transport channel. An isolated Majorana bound state must have $E=0$, otherwise its wave function would not be real. It is this midgap alignment of Majorana bound states that allows for resonant conduction if the density of defects is sufficiently large. The disorder-driven phase transition from a thermal insulator to a thermal metal has not yet been observed experimentally, but it is evident in computer simulations, see Fig.\ \ref{fig_sigmaMI}.\medskip

\begin{figure}[tb]
\centerline{\includegraphics[width=1\linewidth]{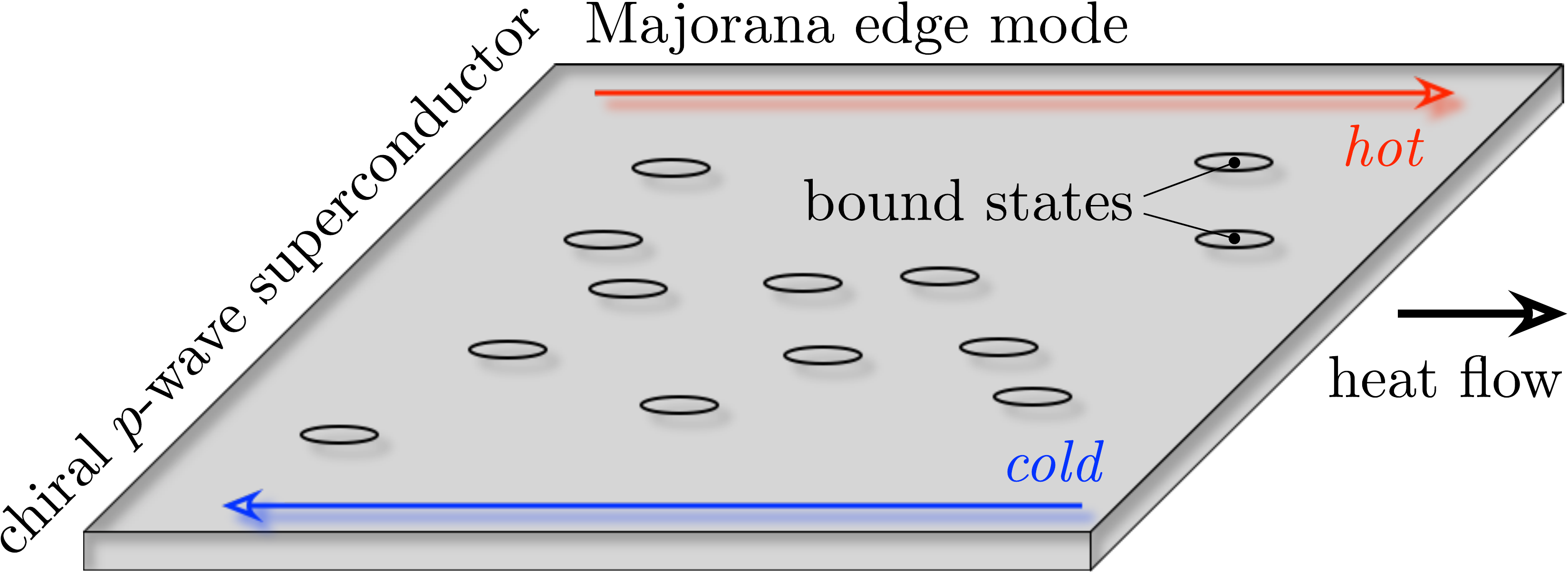}}
\caption{Thermal quantum Hall effect in a 2D topological superconductor: A transverse temperature difference drives a longitudinal heat current, carried by chiral Majorana edge modes. This heat conduction mechanism dominates at low disorder strengths, when the Majorana bound states in the interior are sufficiently far apart that the system has not yet reached the Majorana metal phase of Fig.\ \ref{fig_sigmaMI}.
}
\label{fig_thermalQHE}
\end{figure}

\noindent
\textbf{Thermal quantum Hall effect}\\
In the thermal insulating phase, the Majorana bound states are too far apart to allow for heat conduction in the interior of the system. What remains possible is conduction along the edge. The Majorana edge modes of a chiral \textit{p}-wave superconductor produce the thermal analogue of the quantum Hall effect, see Fig.\ \ref{fig_thermalQHE}. 

We recall that the quantum Hall effect in a semiconductor 2D electron gas is associated both with a quantized electrical conductance and with a quantized thermal conductance. The quantization units are $e^2/h$ and ${\cal L}Te^2/h$, respectively, with $T$ the temperature and ${\cal L}=\tfrac{1}{3}(\pi k_{\rm B}/e)^2$ the Lorenz number. The superconducting counterpart is called the thermal quantum Hall effect, because only the thermal conductance is quantized. The fact that the wave function of a Majorana fermion is real rather than complex, reduces the quantization unit by a factor of two \cite{Sen00}: An unpaired Majorana mode has a thermal conductance of $G_0=\frac{1}{2}{\cal L}Te^2/h$.

The complexity of heat measurements at low temperatures is an obstacle to the detection of the thermal quantum Hall effect, but there is a purely electrical alternative \cite{Gne15}.  While the Majorana edge mode carries no charge on average, it is not in an eigenstate of charge, so there are quantum fluctuations. These produce a quantized shot noise power of $\frac{1}{2}e^2/h$ per eV of voltage bias, where the factor $1/2$ has the same origin as in the thermal conductance quantum.\medskip

\noindent
\textbf{Majorana qubits}\\
While widely separated Majorana bound states are not useful for transport properties, they promise to be very useful for storage of quantum information \cite{Kit01}. Because they are all pinned to $E=0$, they introduce a degeneracy in the ground state of the topological superconductor. The degeneracy factor $2^{N}$ is exponential in the number $N$ of pairs of bound states (``Majorana qubits''), so a massive amount of information can be stored in the ground state. The same information can be stored in quantum superpositions of the states of $N$ electron spins, but such superpositions suffer from dephasing. An isolated Majorana has no phase, hence as long as the bound states remain far apart the quantum information should be protected from dephasing.

An elementary operation on the Majorana qubits is the pairwise exchange (``braiding'') of two Majoranas. If the operation is carried out very slowly, adiabatically, the superconductor remains in the ground state. For a nondegenerate state this would amount to multiplication by a phase factor, but a degenerate state is transformed by a unitary operation. This is the celebrated non-Abelian exchange statistics of Majoranas \cite{Rea00}. (Non-Abelian because unitary operations do not commute.) Not all unitary operations can be obtained by exchanging Majorana qubits, but a hybrid design \cite{Has11,Aas15} that includes also some well-developed superconducting electronics \cite{Lar15,Lan15} provides a road map to a fault-tolerant quantum computer.\medskip

\noindent
\textbf{From 2D to 3D}\\
The central new insight of topological insulators is that topologically nontrivial bandstructures are not limited to 2D systems, such as the quantum Hall insulator: A 3D bulk insulator can have an electrically conducting surface if time-reversal symmetry is not broken \cite{Has10,Qi11}. This insight carries over to topological superconductors: A 3D superconductor can be thermally insulating in the bulk with a thermally conducting surface. The surface conduction is topologically protected in the absence of a magnetic field or magnetic impurities. A promising route to 3D topological superconductivity, followed in Cu-doped Bi$_2$Se$_3$ \cite{Hor10,Sas11}, is to start from a 3D topological insulator and dope it to induce a transition into a superconducting state. The transition brings some remarkable new physics into play. 

The quasiparticles on the surface of a 3D topological insulator are massless Dirac fermions, familiar from graphene. The superconducting counterpart has massless Majorana fermions on its surface. Both quasiparticles have the same relativistic band structure, $E^2=v^2 (p_x^2+p_z^2)$, with energy-independent velocity $v$ and momentum $(p_x,p_z)$ in the $x$-$z$ plane. The Dirac Hamiltonian $H_0=-i\hbar v(\sigma_x\partial/\partial x+\sigma_z\partial/\partial z)$ that produces this band structure (with Pauli matrices $\sigma_x,\sigma_z$) is purely imaginary. For Dirac fermions we may add a disorder potential $V(x,z)$, but this is forbidden for Majorana fermions because $H_0+V$ is then no longer imaginary and the real wave equation would become complex. The physical implication is that Majorana fermions transport heat ballistically over the surface of the topological superconductor, unscattered by disorder. This is a fundamental difference with topological insulators, where disorder cannot localize the surface electrons, but it does scatter them and degrades the ballistic motion to diffusion.\medskip

\noindent
\textbf{Outlook}\\
Thinking ahead about applications of topological insulators one looks at spintronics, because of the spin-momentum locking of the conducting surface electrons. The same ``helicity'' applies to Majorana fermions, but their charge neutrality makes applications in that context less natural. Much of the present research aims at the integration of topological superconductivity into superconducting electronics, with the aim to improve the robustness of a quantum computation by storing information in Majorana bound states. Mobile Majorana fermions, either in edge states or in surface states, have thermal conduction properties that may or may not find applications. What is evident at this time is that topological superconductors provide a laboratory for the study of the remarkable complexity of quantum mechanics without complex numbers.

\acknowledgments
Carlo Beenakker is at the Instituut-Lorentz of Leiden University, Leo Kouwenhoven is at QuTech and the Kavli Institute of Nanoscience of Delft University of Technology.\\
Our collaboration is supported by an ERC Synergy grant and by the NWO/OCW Nanofront consortium.

\end{document}